\documentclass[aps,pra,showpacs,twocolumn,superscriptaddress]{revtex4}
\usepackage{graphicx,color}
\usepackage{amsmath,amsthm,amsfonts,amssymb,bm}
\usepackage{times}
\usepackage{epsf}
\usepackage[colorlinks={true}]{hyperref}

\usepackage[T1]{fontenc}
\usepackage[utf8]{inputenc}

\hypersetup{citecolor={blue}, filecolor={blue}, linkcolor={blue}, urlcolor={blue}}

\newcommand{\commentold}[1]{}
\DeclareMathSymbol{:}{\mathpunct}{operators}{"3A}

\begin{document}

\title{Noisy Metrology: A saturable lower bound on quantum Fisher information}
\author{R. Yousefjani}
\author{S. Salimi}
\email{shsalimi@uok.ac.ir}
\author{A. S. Khorashad}

\affiliation{Department of Physics, University of Kurdistan, P.O.Box 66177-15175 , Sanandaj, Iran}

\date{\today}

\begin{abstract}
In order to provide a guaranteed precision and a more accurate judgement about the true value of the  Cram\'{e}r-Rao bound and its scaling behavior, an
upper bound (equivalently a lower bound on the quantum Fisher information) for precision of estimation is introduced. Unlike the bounds previously introduced in the literature, the
upper bound is saturable and yields a practical instruction to estimate the parameter through preparing the optimal initial state and optimal measurement. The bound is based on the underling
dynamics and its calculation is straightforward and requires only the matrix representation of the quantum maps responsible for encoding the parameter. This allows us to apply the bound to
open quantum systems whose dynamics are described by either semigroup or non-semigroup maps. Reliability and efficiency of the method to predict the ultimate precision limit are demonstrated by {three} main examples.
\end{abstract}


\maketitle

\section{Introduction}\label{I}

Parameters estimation is a principal part of the scientific analysis of experimental data.
It plays an important role at a very fundamental level, involving the measurement of fundamental constants of Nature like the Planck constant, the speed of light in vacuum and the gravitational constant.
Furthermore, it has widespread practical implications ranging from determination of atomic transition frequency \cite{Giovannetti,Wineland,Bollinger} to a phase shift in an interferometric measurement due to the presence of gravitational waves \cite{LIGO,Resch,Higgins}.

Since errors and statistical uncertainties are unavoidable in realistic experimental data, specifying the estimation error is a central task in parameter estimation.
The error in an estimation is quantified by the square root of the statistical average of the squared differences between the true and the estimated values of the parameter.
It is lower bounded by the Cram\'{e}r-Rao bound which, in turn, is inversely proportional to the quantum Fisher information (QFI) in quantum metrology \cite{Book,Braunstein}.
In viewpoint of the information theory, the QFI gives the amount of information about an unknown parameter which can be extracted from scientific analysis of experimental data.
{So, it can be used to characterize the statistical distinguishability of states which are dependent on the parameter and, hence, to indicate the non-Markovian behavior \cite{QFIflow}.}
{It should be point that, the QFI depends on the probe characteristics, the type of the parameter encoding process and the measurement.}
Calculating this quantity, finding ways to maximize it, and designing protocols which allow for better estimation, are central to the quantum metrology.

For a general probe state under a general encoding process (specially noisy process),  due to growing the size of the evolved state exponentially with the number $N$ of the probes, it is a difficult task to compute the QFI and maximize it with respect to the initial probe state and measurement.
To solve this challenge, in addition to some efforts to calculate the maximal QFI in special cases \cite{Knysh,Frowis}, some alternative frameworks have been proposed by deriving fundamental metrological bounds.
Regarding the Kraus representations of the parameter encoding map, some upper bounds on the QFI have been defined \cite{Demk,Escher}.
These bounds are not necessarily tight even after optimization over the equivalent Kraus operators which require numerical methods.
Furthermore, this bound cannot be reached by any measurement strategy.
A more accurate judgment about the true maximal value of the QFI in addition to an upper bound needs a tight and saturable lower bound.
Moreover, to predict the scaling behavior of the QFI, considering the behavior of its lower bound (which gives a guaranteed precision) is more reliable than that of its upper bounds.

Motivated by this, here, a saturable and tight-fitting lower bound on the QFI in open quantum systems under general dynamics (either semigroup or non-semigroup maps) is presented.
The bound is directly related to the underlying dynamics and its calculation requires only the matrix representation of the parameter encoding map which is obtained through tomography process.
By focusing on the frequency estimation in the presence of phase-covariant noise, as the first main example to demonstrate the reliability and efficiency of the method, a useful prescription for practical estimation is provided, by determining the optimal initial state and the optimal measurement.
Surprisingly, the method shows competency to offer the exact precision limit and the optimal initial state in open quantum systems under correlated noise (dephasing), as the second example.
{In the last example, the performance of the bound in predicting the optimal initial subspace for a phase estimation in an optical interferometry with photon loss is briefly discussed.}
{Note that, in a variant approach, Alipour et al. \cite{Rezakhani1} and Beau et al. \cite{Beau}, have presented some bounds on the QFI for determined initial state in open quantum systems only with dynamical semigroup maps.}

\section{Bound on the quantum Fisher information}
In measuring a parameter $x$, the uncertainty $\delta x$ with which the parameter can be measured is lower bounded by the Cram\'{e}r-Rao bound  \cite{Book,Braunstein}
\begin{equation}\label{Uncertanity}
 (\delta x)^{2} \geq \frac{1}{\frac{T}{t} \max_{\rho_{0}}\mathcal{F}(\rho_{x})},
\end{equation}
where $t$ is the duration of each single measurement on the $N$ probes and $\frac{T}{t}$ is the times that one repeats the measurement in a given fixed time $T$.
$\mathcal{F}(\rho_{x})$ is the QFI which is a measure of the amount of information in the encoded state $\rho_{x}$ about the parameter $x$ and can be maximized over different initial states.
Due to the convexity of the QFI \cite{Fujiwara}, the optimal initial states are pure.
From mathematical point of view, the QFI is defined as
$\mathcal{F}(\rho_{x})= tr(\rho_{x} \mathrm{L}_{x}^{2})$ where $\mathrm{L}_{x}$ is a symmetric logarithmic derivative operator which satisfies the equation $\partial_{x}\rho_{x}=\frac{1}{2}(\rho_{x}\mathrm{L}_{x}+\mathrm{L}_{x}\rho_{x})$.
If the eigenvalues and eigenvectors of $\rho_{x}$ are known, the calculation of the QFI will be an easy task \cite{Braunstein}.
However, very often the analytical diagonalization of $\rho_{x}$ turns out not to be feasible.

On the other hand, the QFI is naturally related to distinguishability of the states in the manifold of quantum states
and is proportional to the Bures distance \cite{Braunstein,Bures} between $\rho_{x}$ and its neighbor states $\rho_{x+dx}$ via
$\mathcal{F}(\rho_{x}) dx^{2}=4 d_{B}^{2}(\rho_{x},\rho_{x+dx})$, where
\begin{equation}
d_{B}^{2}(\rho_{x},\rho_{x+dx})= 2\left(1-tr\sqrt{\sqrt{\rho_{x}}\rho_{x+dx}\sqrt{\rho_{x}}}\right).
\end{equation}
Therefore, one will exploit the above relation to access the analytical formula of QFI if one obtains the explicit expression of the Bures distance which is difficult to obtain.
Here, this difficulty is circumvented by studying the statistics of an open quantum system through the behavior of states in the Liouville space, $\mathcal{L}(\mathcal{H})$, which is the vector space formed by the set of all linear operators acting on the Hilbert space $\mathcal{H}$.
The Liouville vectors, $\{|\mathrm{A})\}$, which correspond to the operators $\{\mathrm{A}\}$ acting on $\mathcal{H}$, satisfy an inner product as $(\mathrm{A}|\mathrm{B})= tr[\mathrm{A}^{\dag}\mathrm{B}]$.
Let  $\Phi$ be a linear map defined on $\mathcal{L}(\mathcal{H})$ which relates one Liouville vector to another as $\tilde{\Phi}|\mathrm{A})=|\Phi \mathrm{A})$, where $\tilde{\Phi}$ is the matrix representation of $\Phi$.
The one-to-one correspondence between $\Phi$ and $\tilde{\Phi}$ is induced by the inner product.
On this basis, any state such as $\rho_{x}$ on the Hilbert space $\mathcal{H}$ has a corresponding vector $|\rho_{x})$ in $\mathcal{L}(\mathcal{H})$.
Now, by defining a normalized pure state as $\tilde{\rho}_{x}=|\Psi_{x})(\Psi_{x}|$ with $|\Psi_{x})=|\rho_{x})/\sqrt{(\rho_{x}|\rho_{x})}$, the Bures distance between $\tilde{\rho}_{x}$ and $\tilde{\rho}_{x+dx}$ is written as
\begin{equation}\label{1}
\tilde{d}_{B}^{2}(\tilde{\rho}_{x},\tilde{\rho}_{x+dx})=2\left(1-\sqrt{(\Psi_{x}|\tilde{\rho}_{x+dx}|\Psi_{x})}\right).
\end{equation}
Considering the Taylor series of $\tilde{\rho}_{x+dx}$ around $x$ and ignoring the third and higher powers of $dx$, Eq. (\ref{1}) reduces to
\begin{eqnarray}\label{Bures distance}
 \nonumber \tilde{d}_{B}^{2}(\tilde{\rho}_{x},\tilde{\rho}_{x+dx}) &=&\frac{(\rho^{\prime}_{x}|\rho^{\prime}_{x})(\rho_{x}|\rho_{x})-(\rho^{\prime}_{x}|\rho_{x})(\rho_{x}|\rho^{\prime}_{x})}{(\rho_{x}|\rho_{x})^{2}}dx^{2} \\
   &=&\frac{1}{4} \tilde{\mathcal{F}}(\tilde{\rho}_{x})dx^{2},
\end{eqnarray}
where $\rho^{\prime}_{x}=\partial_{x}\rho_{x}$ and $\tilde{\mathcal{F}}(\tilde{\rho}_{x})=tr(\tilde{\rho}_{x}\tilde{L}_{x}^{2})$ is an associated QFI which quantifies the amount of information in $\tilde{\rho}_{x}$ and $\tilde{L}_{x}=2\partial_{x}\tilde{\rho}_{x}$.
To find the relation between  $\mathcal{F}(\rho_{x})$  and $\tilde{\mathcal{F}}(\tilde{\rho}_{x})$, one should use the definition of the QFI in terms the symmetric logarithmic derivative  operator, $\mathrm{L}_{x}$.
Replacing $\rho^{\prime}_{x}$ by $\frac{1}{2}(\rho_{x}\mathrm{L}_{x}+\mathrm{L}_{x}\rho_{x})$ in the first term of $\tilde{\mathcal{F}}(\tilde{\rho}_{x})$ results in
\begin{equation}\label{s2}
 \tilde{\mathcal{F}}(\tilde{\rho}_{x}) = \frac{4}{(\rho_{x}|\rho_{x})}\frac{1}{2}(tr(\mathrm{L}_{x}\rho_{x}\mathrm{L}_{x}\rho_{x})+tr(\mathrm{L}_{x}^{2}\rho_{x}^{2}))-\chi,
\end{equation}
where $\chi=4(\rho^{\prime}_{x}|\rho_{x})(\rho_{x}|\rho^{\prime}_{x})/(\rho_{x}|\rho_{x})^{2}$.
Employing the positivity of $\rho_{x}$, $\mathrm{L}_{x}\rho_{x}\mathrm{L}_{x}$, and $\sqrt{\rho_{x}}\mathrm{L}_{x}^{2}\sqrt{\rho_{x}}$, we obtain
\begin{equation}\label{s3}
tr(\mathrm{L}_{x}\rho_{x}\mathrm{L}_{x}\rho_{x})+tr(\sqrt{\rho_{x}}\mathrm{L}_{x}^{2}\sqrt{\rho_{x}}\rho_{x})\leq 2tr(\mathrm{L}_{x}^{2}\rho_{x})=2\mathcal{F}(\rho_{x}).
\end{equation}
By combining Eq. (\ref{s2}) and (\ref{s3}), one will have
\begin{equation}
\tilde{\mathcal{F}}(\tilde{\rho}_{x}) \leq \frac{4}{(\rho_{x}|\rho_{x})}\mathcal{F}(\rho_{x})-\chi,
\end{equation}
which, in turn, can be rearranged as
\begin{equation}
\frac{(\rho_{x}|\rho_{x})}{4}\tilde{\mathcal{F}}(\tilde{\rho}_{x})\leq\frac{(\rho_{x}|\rho_{x})}{4}(\tilde{\mathcal{F}}(\tilde{\rho}_{x})+\chi) \leq \mathcal{F}(\rho_{x}).
\end{equation}
The left hand side of this equation which can be simplified to
\begin{equation}\label{bound}
F^{\downarrow}(\rho_{x})=(\rho^{\prime}_{x}|\rho^{\prime}_{x})-\frac{(\rho^{\prime}_{x}|\rho_{x})(\rho_{x}|\rho^{\prime}_{x})}{(\rho_{x}|\rho_{x})},
\end{equation}
is our lower bound of the QFI.
The optimal measurement which saturates this lower bound uses  positive operator-valued measures $\{E_{j}\}$  which are one-dimensional projection operators onto the non-degenerate eigenspace of the Hermitian operator $\tilde{L}_{x}$.

In a typical estimation setting, the parameter $x$ is encoded on sensing probes by a given physical dynamics $\Phi_{x}$ with $\tilde{\Phi}_{x}$ as the corresponding matrix representation.
So, Eq. (\ref{bound}) can be rewritten as
\begin{eqnarray}\label{6}
F^{\downarrow}(\rho_{x}) &=&  \{(\rho_{0}|\tilde{\Phi}_{x}^{\prime \dag}\tilde{\Phi}_{x}^{\prime}|\rho_{0})-\frac{|(\rho_{0}|\tilde{\Phi}_{x}^{\prime \dag}\tilde{\Phi}_{x}|\rho_{0})|^{2}}{(\rho_{0}|\tilde{\Phi}_{x}^{\dag}\tilde{\Phi}_{x}|\rho_{0})}\}.
\end{eqnarray}
As can be seen, the bound is directly related to the underlying dynamics and its determination only needs the knowledge of the matrix representation of the parameter encoding map.
This property allows one to use the bound in open quantum systems which are governed by either semigroup or non-semigroup dynamical maps.

In particular, for a closed system evolving under a unitary transformation, $\mathcal{U}_{x}$, and being prepared in initial pure state $\rho_{0}$, one can show that $(\rho_{x}|\rho^{\prime}_{x})=0$ and the bound reduces to
\begin{equation}\label{bound for pure state}
F^{\downarrow}=(\rho_{0}|\tilde{\mathcal{U}}_{x}^{\prime \dag}\tilde{\mathcal{U}}_{x}^{\prime}|\rho_{0})=\frac{1}{2}\mathcal{F}(\rho_{x}).
\end{equation}
However, $\max_{\rho_{0}}(\rho_{0}|\tilde{\mathcal{U}}_{x}^{\prime \dag}\tilde{\mathcal{U}}_{x}^{\prime}|\rho_{0})\leqslant \mathcal{F}(\rho_{x})$
does hold in general, if the maximization were not restricted to physical state, but extended to all normalized  vectors in Liouville space, one would end up calculating the operator norm $\|\tilde{\mathcal{U}}_{x}^{\prime \dag}\tilde{\mathcal{U}}_{x}^{\prime}\|$, which equals to the largest eigenvalue of the enclosed matrix.
For the closed system, this quantity yields $\mathcal{F}(\rho_{x})$ and ,as will be illustrated bellow, the subspace of the optimal initial state is determined by the components of the associated eigenstate.
Since, one deals only with matrix representation of the encoding map, our framework can be used to randomly sampled generators, as is provided by  Nichols et al. \cite{Adesso}.

Furthermore, while $\tilde{\mathcal{F}}(\tilde{\rho}_{x})$ is an additive function, i.e.,
$\tilde{\mathcal{F}}(\tilde{\rho}_{x}^{\otimes \nu})=\nu\tilde{\mathcal{F}}(\tilde{\rho}_{x})$,
the lower bound is a subadditive function,
$F^{\downarrow}(\rho_{x}^{\otimes \nu})\leq \nu F^{\downarrow}(\rho_{x})$.

In the following, the attention is paid to the utility of the bound in predicting correct behavior (e.g., scaling) of the estimation error in an atomic spectroscopy (or equivalently magnetic field sensing) in the presence of uncorrelated and correlated noise.
Derived results are widely applicable to a broad range of relevant physical processes including  noisy depolarization, such as spin-lattice relaxation at room temperature.
{Moreover, we complete the paper by a brief discussion on the performance of the bound for phase estimation in a lossy interferometry.}


\section{Frequency estimation}
In a frequency estimation scenario, a parameter $\omega$ will be encoded on each qubit 
by a unitary encoding map as
$\mathcal{U}_{\omega}[O]=e^{\frac{-i\omega t}{2}\sigma_{z}}Oe^{\frac{i\omega t}{2}\sigma_{z}}$, where $\sigma_{z}$ is the Pauli operator generating a rotation of the qubit state around the $z$-axis in the Bloch ball representation.
Selecting the computational basis as $\{|\mu\nu)\}$ in $\mathcal{L}(\mathcal{H})$
($\{\mu\nu =|\mu\rangle\langle\nu|\}$ with $\mu,\nu\in\{0,1\}$)
the matrix representation of ${\mathcal{U}}_{\omega}$ can be obtained as
$\tilde{\mathcal{U}}_{\omega}=\sum _{\mu\nu}e^{i\alpha \omega t}|\mu\nu)(\mu\nu|$,
where $\alpha=\mu-\nu$.
In the case of $N$ identical qubits, the matrix representation of $\mathcal{U}_{\omega}^{\otimes N}$ can be obtained by the matrix product $\tilde{\mathcal{U}}_{\omega}^{\otimes N}$ which is a $2^{2N}\times2^{2N}$ diagonal matrix whose elements are $e^{i\alpha_{_{\emph{N}}}\omega t }$, where $-N\leq\alpha_{\emph{N}}\leq N$.
Recall from the previous discussions that
\begin{equation}\label{bound in close system}
F^{\downarrow}\leq\|(\tilde{\mathcal{U}}_{\omega}^{\otimes N})^{\prime \dag} (\tilde{\mathcal{U}}_{\omega}^{\otimes N})^{\prime}\|= N^{2}t^{2}.
\end{equation}
Associated eigenstate leads one to preparing the initial state in the Greenberger-Horne-Zeilinger (GHZ) form (see Appendix A).

\subsection{Frequency estimation in the presence of uncorrelated phase-covariant noise}
In this case each qubit is locally affected by a special noise type named as phase-covariant noise, that is, a noise type commuting with the parameter encoding Hamiltonian, $\sigma_z$, \cite{Demk2,Holevo}.
In the case of semigroup dynamics, the noise type is one of the most destructive noise due to constraining the quantum enhancement to a constant factor \cite{Demk,Huelga,Rezakhani1}.
However, this is not the case for a non-semigroup dynamics \cite{Demk2,Matsuzaki,Chin,K,Me}.
Regarding $\mathcal{U}_{\omega}$ as the encoding unitary map and $\mathcal{J}$ as the parameter-independent noise map, the state of $N$ probes at any instant is described as
\begin{equation}\label{2}
\rho_{\omega}=\Phi_{\omega}^{\otimes N}[\rho_{0}],
\end{equation}
with
\begin{equation}
\Phi_{\omega}=\mathcal{U}_{\omega}\circ \mathcal{J}=\mathcal{J}\circ \mathcal{U}_{\omega}.
\end{equation}
Considering the most general form of the phase-covariant noise, the matrix form of
$\Phi_{\omega}$ in the computational basis is obtained as \cite{Demk2}
\begin{equation}\label{covariant map1}
\tilde{\Phi}_{\omega} = \begin{pmatrix} J_{++} & 0 & 0 & J_{+-}\\ 0 & 0 & \eta_{\perp}e^{-i\phi} & 0\\ 0 & \eta_{\perp}e^{i\phi} & 0 & 0 \\ J_{--} & 0 & 0 & J_{-+} \end{pmatrix},
\end{equation}
where $J_{\pm\pm}=\frac{1\pm k\pm\eta_{\parallel}}{2}$.
The map includes a rotation around the $z$-axis by an angle $\phi$ containing the encoded parameter, $\omega$, as $\phi=\omega t+\theta$, a symmetric contraction in the $xy$ plane by a factor $0\leq\eta_{\perp}\leq1$, a contraction in the $z$ direction by a factor $-1\leq\eta_{\parallel}\leq 1$ ( the case  $\eta_{\parallel}\leq 0$ corresponds to an additional reflection with respect to the $xy$ plane), and a displacement in the $z$ direction by $-1\leq k\leq 1$.
The map in Eq. (\ref{covariant map1}) fulfils the completely positive and trace preserving conditions as long as $\eta_{\parallel}\pm k\leq 1$ and $1+\eta_{\parallel}\geq\sqrt{k^{2}+4\eta_{\perp}^{2}}$ \cite{Demk2}.
In particular, by setting special instances of the noise parameters, typical qubit channels like pure dephasing
($k=0$, $\eta_{\parallel}=1$ and $\eta_{\perp}>0$), isotropic depolarisation ($k=0$ and $\eta_{\parallel}=\eta_{\perp}>0$) and amplitude damping
($\eta_{\parallel}=1-k$ and $\eta_{\perp}=\sqrt{1-k}$) can be obtained.
The following theorem provides an alternative way of evaluating the maximum of $F^{\downarrow}$ in the presence of the phase-covariant noise.

\textit{Theorem: Let $\Phi_{\omega}$ be the encoding map in the presence of a phase-covariant noise. Then}
\begin{equation}\label{theorem}
F^{\downarrow}\leq\|(\tilde{\Phi}_{\omega}^{\otimes N})^{\prime \dag} (\tilde{\Phi}_{\omega}^{\otimes N})^{\prime}\|,
\end{equation}
\textit{and initially preparing qubits in the GHZ state results in
$F^{\downarrow}=\frac{1}{2}\|(\tilde{\Phi}_{\omega}^{\otimes N})^{\prime \dag} (\tilde{\Phi}_{\omega}^{\otimes N})^{\prime}\|$.}

\textit{Proof:} Since $\mathcal{U}_{\omega}$ and $\mathcal{J}$ commute, one can have
$|\rho_{\omega}^{\prime})=(\tilde{\mathcal{U}}_{\omega}^{\otimes N})^{\prime}(\tilde{\mathcal{J}}^{\otimes N})|\rho_{0})$.
Therefore, one obtains  $(\rho_{\omega}^{\prime}|\rho_{\omega})=(\rho_{0}|(\tilde{\mathcal{J}}^{\otimes N})^{\dagger}(\tilde{\mathcal{U}}_{\omega}^{\otimes N})^{\prime\dagger}(\tilde{\mathcal{U}}_{\omega}^{\otimes N})(\tilde{\mathcal{J}}^{\otimes N})|\rho_{0})=0$.
This also occurs when the parameter independent noise do not commute with $\mathcal{U}_{\omega}$ but can be suppressed after the sensing transformation, that is, states to be measured are $\tilde{\mathcal{U}}_{\omega}^{\otimes N}|\varrho)$ with $|\varrho)=\tilde{\mathcal{J}}^{\otimes N}|\rho_{0})$.
As discussed in the Appendix A, the largest eigenvalue of
$(\tilde{\mathcal{U}}_{\omega}^{\otimes N})^{\prime\dag}(\tilde{\mathcal{U}}_{\omega}^{\otimes N})^{\prime}$ is $N^{2}t^{2}$ with the corresponding eigenvectors $|01^{\otimes N})$ and $|10^{\otimes N})$ which under the action of $(\tilde{\mathcal{J}}^{\otimes N})$
are changed to $(\eta_{\perp}e^{-i\theta})^{N}|10^{\otimes N})$ and $(\eta_{\perp}e^{i\theta})^{N}|01^{\otimes N})$, respectively.
This shows that the eigenspace corresponding to the largest eigenvalue of
$(\tilde{\mathcal{U}}_{\omega}^{\otimes N})^{\prime\dag}(\tilde{\mathcal{U}}_{\omega}^{\otimes N})^{\prime}$ is invariant under the phase-covariant noise.
This also occurs for the eigenspace corresponding to the smallest eigenvalue of
$(\tilde{\mathcal{U}}_{\omega}^{\otimes N})^{\prime\dag}(\tilde{\mathcal{U}}_{\omega}^{\otimes N})^{\prime}$ (i.e., zero), which is spanned by $(|00)+|11))^{\otimes N}$, due to
\begin{eqnarray}
\nonumber \tilde{\mathcal{J}}|00) &=& J_{++} |00)+J_{--} |11),\\
 \tilde{\mathcal{J}}|11) &=& J_{+-} |00)+J_{-+} |11),\\
 \nonumber
\end{eqnarray}
where $J_{\pm\pm}=\frac{1\pm k\pm\eta_{\parallel}}{2}$.
So, preparing the initial state as $|\rho_{0})=\frac{|u_{\textit{max}})+|u_{\textit{min}})}{\sqrt{2}}$ with
$|u_{\textit{max}})=\frac{|01^{\otimes N})+|10^{\otimes N})}{\sqrt{2}}$ and $|u_{\textit{min}})=\frac{|00^{\otimes N})+|11^{\otimes N})}{\sqrt{2}}$
provides us with $F^{\downarrow}=\frac{1}{2}\|(\tilde{\Phi}_{\omega}^{\otimes N})^{\prime \dag} (\tilde{\Phi}_{\omega}^{\otimes N})^{\prime}\|$.
 $\qquad \qquad \qquad \qquad \quad \quad \qquad \quad \quad $  $\square$

The theorem shows that as long as $0\leq t<\tau$, the maximum of the lower bound grows quadratically for small $t$ and  is fully determined by $\eta_{\perp}$ in the plane perpendicular to the rotation axis, as (Appendix B)
\begin{eqnarray}\label{3}
\|(\tilde{\Phi}_{\omega}^{\otimes N})^{\prime \dag} (\tilde{\Phi}_{\omega}^{\otimes N})^{\prime}\| = N^{2}t^{2}\eta_{\perp}^{2N},
\end{eqnarray}
and $\tau$ is the largest time  satisfying
\begin{equation}
\frac{2N^{2}}{(N-1)^{2}}\eta_{\perp}^{2}=1+k^{2}+\eta_{\parallel}^{2}+\sqrt{(1+k^{2}+\eta_{\parallel}^{2})^{2}-4\eta_{\parallel}^{2}},
\end{equation}
for the short-time expansion of the noise parameters,
\begin{eqnarray}
\nonumber \eta_{\perp} &=& 1-\alpha_{\perp}t^{\beta_{\perp}}+..., \\
\nonumber \eta_{\parallel} &=& 1-\alpha_{\parallel}t^{\beta_{\parallel}}+...,\\
  k &=& \alpha_{k}t^{\beta_{k}}+...\quad .
\end{eqnarray}
In a noisy metrology, after a time, the noise wins over the unitary encoding process and the extractable information is extremely degraded.
So, we need to find the optimal interrogation time, $t_{\textit{opt}}$.
Equivalently, this corresponds to computing ${\min_{t}}(\frac{t}{ \max_{\rho_{0}}F^{\downarrow}})$,  which results in $t_{\textit{opt}}=(2\alpha_{\perp}N(\beta_{\perp}+1))^{-1/\beta_{\perp}}<\tau$.
Therefore, interrogating the probes in interval $0<t\leq t_{\textit{opt}}$ leads to the following upper bound for the Cram\'{e}r-Rao bound
\begin{eqnarray}\label{4}
  \min_{t} \frac{t}{T \max_{\rho_{0}}\mathcal{F}(\rho_{x})} \leq \frac{C^{\downarrow}}{N^{(2\beta_{\perp}-1)/\beta_{\perp}}},
\end{eqnarray}
where $C^{\downarrow}=(2\alpha_{\perp})^{1/\beta_{\perp}}(1+\beta_{\perp})^{(\beta_{\perp}+1)/\beta_{\perp}}/T\beta_{\perp}$.
Crucially, this result proves that the scaling of the uncertainty, $N^{-(2\beta_{\perp}-1)/\beta_{\perp}}$, predicted by the finite-$N$ channel extension method $\frac{C^{\uparrow}}{N^{(2\beta_{\perp}-1)/\beta_{\perp}}}$ \cite{Demk2} is indeed always achievable up to a constant factor.
However, the lower bound of the Cram\'{e}r-Rao bound predicts the ultimate precision limit (which cannot be reached by any measurement strategy), by limiting the Cram\'{e}r-Rao bound from both sides
\begin{figure}[b]
\includegraphics[scale=.92]{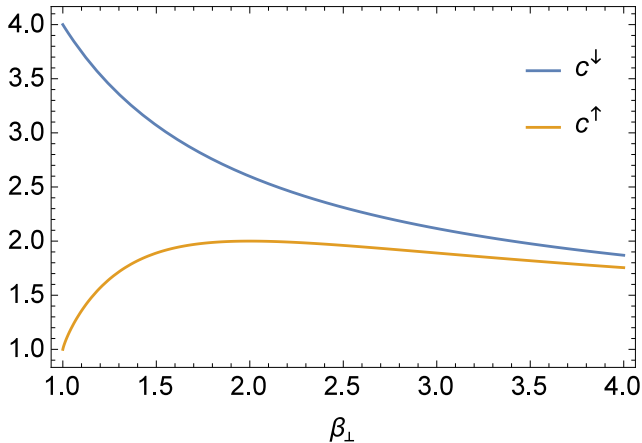}
\caption{(Color online) Dynamics of $C^{\downarrow}$ (up) and $C^{\uparrow}$ (down) for $\beta_{\perp}<\beta_{\|}$ and $\alpha_{\perp}=1/2$.}
\label{fig1}
\end{figure}
\begin{equation}
\frac{C^{\uparrow}}{N^{(2\beta_{\perp}-1)/\beta_{\perp}}}\leq \min_{t} \frac{t}{T \max_{\rho_{0}}\mathcal{F}(\rho_{x})} \leq \frac{C^{\downarrow}}{N^{(2\beta_{\perp}-1)/\beta_{\perp}}},
\end{equation}
one has a more accurate judgment about the true maximal value of the precision (Fig. \ref{fig1}).
While, for semigroup dynamics ($\beta_{\perp}=1$) one accordingly recovers the standard quantum limit scaling, $N^{-1}$, one can find more favourable scaling by going beyond the semigroup regime (e.g., by exploiting the non-semigroup dynamics arising at short times in the Zeno regime, $\beta_{\perp}=2$).

Saturating the quantum Cram\'{e}r-Rao bound is subject to the initial preparation of the probes in an optimal state and choosing the optimal measurement strategy, on one hand, and repeating the above-mentioned actions infinite times ($T\rightarrow \infty$), on the other hand.
Although a general solution has not yet been known for the optimal measurements which saturates the quantum Cram\'{e}r-Rao bound, the sufficient condition for a measurement to be an optimal one is that it projects the state onto the eigenspaces of the symmetric logarithmic derivative operator \cite{Braunstein}.
Similar investigation leads to a sufficient condition for obtaining the optimal measurements providing Eq. (\ref{theorem}) and, therefore, the upper bound of the Cram\'{e}r-Rao bound. This condition restricts the measurements to one-dimensional projectors onto the non-degenerate eigenspaces of $\rho^{\prime}_{x}$ (Appendix C).
The construction yields the measurements which are in principle not only collective, i.e. act on all the particles, but also local in the parameter space.


\subsection{Frequency estimation in the presence of correlated noise}
For closely spaced particles, particulary ions stored in linear Paul traps or atoms in optical lattices, correlated dephasing is a major source of noise \cite{Monz,Roos,La,Ki,Dor}.
Consider a scheme consisting of $N$ identical probes.
Every single probe, in turn, comprises two two-level atoms with different transition frequencies $\omega_{1}$ and $\omega_{2}$.
The frequency difference $\bar{\omega}=\omega_{1}-\omega_{2}$ is going to be estimated by performing the standard Ramsey-type measurement.
In the presence of correlated dephasing, the encoding process changes the state of the $N$ probes to
\begin{equation}
\rho_{\omega_{1},\omega_{2}}=\Phi_{\omega_{1},\omega_{2}}[\rho_{0}].
\end{equation}
Deriving the matrix form of $\Phi_{\omega_{1},\omega_{2}}$ in the computational basis
$\{|\mu_{1}\nu_{1}\otimes \mu_{2}\nu_{2}\otimes..\otimes\mu_{N}\nu_{N})\}$ in $\mathcal{L}(\mathcal{H}^{\otimes N})$  (with $\mu_{1}\nu_{1}\otimes..\otimes\mu_{N}\nu_{N}=|\mu_{1}\rangle\langle\nu_{1}|\otimes..\otimes|\mu_{N}\rangle\langle\nu_{N}|$ and $\mu_{i},\nu_{i}\in\{0,1\}$ for $i=1,..,N$), results in a diagonal matrix with the elements $e^{i(\alpha_{1}\omega_{1}+\alpha_{2}\omega_{2})t-\alpha^{2}\gamma t}$, where
\begin{eqnarray}
\nonumber  \alpha_{1} &=& \sum_{i=1,3,..,2N-1}(\mu_{i}-\nu_{i}), \\
\nonumber  \alpha_{2} &=& \sum_{i=2,4,..,2N}(\mu_{i}-\nu_{i}), \\
  \alpha &=&  \alpha_{1}+ \alpha_{2},
\end{eqnarray}
and $\gamma$ is dephasing rate.
Since the noise is compatible with the encoding Hamiltonian, the maximum of the lower bound is given by Eq. (\ref{theorem}).
After some straightforward computations, one finds the elements of the diagonal matrix
$\tilde{\Phi}^{\prime \dag}_{\omega_{1},\omega_{2}}\tilde{\Phi}^{\prime}_{\omega_{1},\omega_{2}}$ as
$\alpha_{1,2}^{2}t^{2}e^{-2(|\alpha_{1}|-|\alpha_{2}|)\gamma t}$ and $ \alpha_{1,2}^{2}t^{2}$.
These results lead to $\|\tilde{\Phi}^{\prime \dag}_{\omega_{1},\omega_{2}}\tilde{\Phi}^{\prime}_{\omega_{1},\omega_{2}}\|=N^{2}t^{2}$ which predicts that the correlated dephasing may not destroy our frequency measurements and the eigencpace corresponding to the eigenvalue $N^{2}t^{2}$ form a decoherence-free subspace with respect to the noise.
Hence, frequency measurements in the presence of the correlated noise can be done by the Heisenberg precision scaling as was shown in \cite{Dor}.

\section{Phase estimation in a lossy interferometry}

{It is interesting to briefly discus the well known issue of phase estimation in an optical two-arm interferometry in the presence of photon loss in one arm.
It is known that, for this case, among quantum states with a definite photon number, $N$, states as
$\sqrt{p}|m,N-m\rangle+ \sqrt{1-p} |N,0\rangle$ with $m\neq N$ are more beneficial than $N00N$ state (i.e., $\frac{1}{\sqrt{2}}(|N,0\rangle+|0,N\rangle)$) which completely miss their coherence by losing a photon \cite{Dorner}.
In an optical interferometry, loss can be modeled by fictitious beam splitter of transmissivity $\eta$ (ranging between $0$ for complete
losses and $1$ for no losses) on the same arm which accumulate phase shift through  $U_{\varphi}=e^{-i \varphi \widehat{a}^{\dagger}\widehat{a}}$, where $\widehat{a}$ is the annihilation operator for arm $a$.
Since the noise operation and the phase accumulation commute, the maximum of our lower bound is given by the largest eigenvalue of $\tilde{\Phi}_{\varphi}^{\prime \dagger}\tilde{\Phi}_{\varphi}^{\prime}$ which is a diagonal matrix in Fock basis
$\{\bigg ||k,N-k\rangle\langle m,N-m|\bigg)\}_{k,m=0}^{N}$ in Liouville space.
After some straightforward computations, the elements of the diagonal matrix  $\widetilde{\Phi}_{\varphi}^{\prime \dagger}\widetilde{\Phi}_{\varphi}^{\prime}$
can be obtained as
\begin{equation}
\sum_{l=0}^{N}(k-m)^{2}\begin{pmatrix} k \\ l \end{pmatrix}\begin{pmatrix} m \\ l \end{pmatrix} \eta^{k+m-2l}(1-\eta)^{2l},
\end{equation}
where $l$ is associated with the number of lost photons.
Our considerations show that the largest eigenvalue is obtained for $k=N$ and $m\neq N$.
By decreasing the noise parameter, $\eta$, the optimal $m_{\textit{max}}$ increases (see Fig. \ref{fig2}).
\begin{figure}[b]
\includegraphics[scale=.92]{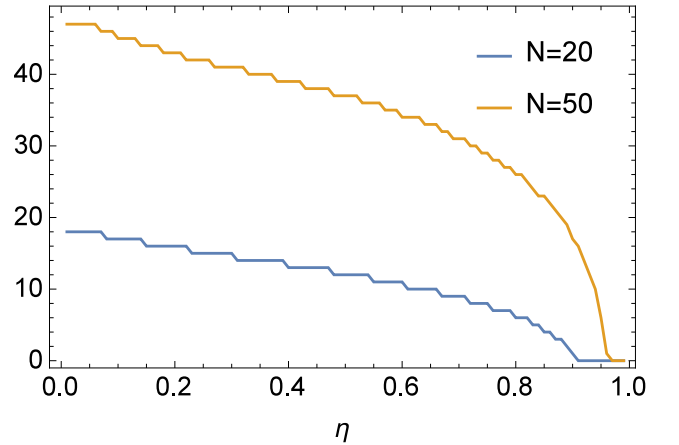}
\caption{(Color online) Dynamics of $m_{\textit{max}}$ versus in two-arm interferometry with $N=20$ (down) and $N=50$ (up) particles.}
\label{fig2}
\end{figure}
This shows that our lower bound can exactly predict the subspace $\{|m,N-m\rangle, |N,0\rangle\}$  as the optimal subspace.}

{Recently the performance of other states such as entangled coherent states, $\mathcal{N}_{\alpha}(|\alpha,0\rangle+|0,\alpha\rangle)$ with $\mathcal{N}_{\alpha}=[2(1+e^{-|\alpha|^{2}})]^{-1/2}$ as normalization constant, for quantum-enhanced phase estimation is investigated \cite{Zhang}.
For these special initial states, our bound leads to $F^{\downarrow}=2\overline{n}\eta f_{C}+ (\overline{n}\eta)^{2} f_{H}$, where
$\overline{n}=\langle \widehat{n}\rangle=2 \mathcal{N}_{\alpha}^{2}|\alpha|^{2}$, $f_{C}=\frac{\mathcal{N}_{\alpha}^{2}}{4} \xi$, $f_{H}=\frac{1}{2}(-1+\xi/2)$ and $\xi=(1+e^{-(1-\eta)|\alpha|^{2}})^{2}(1+e^{-\eta|\alpha|^{2}})+(1-e^{-(1-\eta)|\alpha|^{2}})^{2}(1-e^{-\eta|\alpha|^{2}})$.
Under practical conditions: $\eta\sim 1$ and $|\alpha|^{2}\gg 1$, one has $f_{C}=(1+e^{-2(1-\eta)|\alpha|^{2}})/4$ and $f_{H}=e^{-(1-\eta)|\alpha|^{2}}/2$.
When the number of photons being lost $(1-\eta)|\alpha|^{2}\ll 1$, the Heisenberg term,
$F^{\downarrow}\approx (\overline{n}\eta)^{2}e^{-(1-\eta)|\alpha|^{2}}$, dominates.
With the increase of $(1-\eta)|\alpha|^{2}$, the classical term, $\overline{n}\eta$, becomes important.
So, even for such special initial states, the bound leads to the  same results of \cite{Zhang} both in terms of scaling and in terms of dominant behavior.}

\section{Conclusions}

Specifying the effect of noise on the ultimate precision limit is a crucial element in developing quantum techniques for metrological tasks.
However, determination of the ultimate precision limit which is given by the Cram\'{e}r-Rao bound in noisy metrology becomes more and more cumbersome when the number of resources increases.
{Although}, some  previously derived lower bounds on the precision delimits the Cram\'{e}r-Rao bound from bellow, a more accurate judgment about the true maximal value and the scaling behavior of the precision needs an upper bound which give a guaranteed precision.
Here, a reliable and saturable lower bound on the QFI in a single-parameter estimation has been introduced to provide this necessity.
This bound provides us with a guaranteed precision and allows us to estimate the ultimate precision limit with an acceptable accuracy.
It has been shown that the lower bound depends only on the underlying dynamics and its calculation requires only the matrix representation of the parameter encoding map.
This property allows one to use the bound in open quantum systems which are governed by either semigroup or non-semigroup dynamical maps.
Moreover, unlike the previously introduced bounds on the QFI, determining the optimal probe state and optimal measurement in our framework suggests a useful prescription for practical frequency estimation.
{The accuracy and efficiency of our method to predict the ultimate precision limit and the optimal initial state have been illustrated through three main examples: frequency estimation in the presence of uncorrelated and correlated noise and phase estimation in a lossy interferometry.}

\textbf{Acknowledgment}: We would like to thanks V. Karimipour and L. Maccone, for helpful discussions.


\appendix

\section{Derivation of Eq. (\ref{bound in close system})}

In the case of $N$ identical qubits, one can consider the basis $\{|\mu_{1}\nu_{1}\otimes \mu_{2}\nu_{2}\otimes..\otimes\mu_{N}\nu_{N})\}$ in $\mathcal{L}(\mathcal{H}^{\otimes N})$ (with $\mu_{1}\nu_{1}\otimes..\otimes\mu_{N}\nu_{N}=|\mu_{1}\rangle\langle\nu_{1}|\otimes..\otimes|\mu_{N}\rangle\langle\nu_{N}|$ and $\mu_{i},\nu_{i}\in\{0,1\}$ for $i=1,..,N$).
In this basis

\begin{equation}
\tilde{\mathcal{U}}_{\omega}^{\otimes N}=\sum e^{i\alpha_{_{\emph{N}}}\omega t}|\mu_{1}\nu_{1}\otimes..\otimes\mu_{N}\nu_{N})(\mu_{1}\nu_{1}\otimes ..\otimes\mu_{N}\nu_{N}|,
\end{equation}
where $\alpha_{_{\emph{N}}}=\sum_{i=1}^{N}(\mu_{i}-\nu_{i})$ which leads to $-N\leq\alpha_{_{\emph{N}}}\leq N$.
Calculation of $(\tilde{\mathcal{U}}_{\omega}^{\otimes N})^{\prime\dag}(\tilde{\mathcal{U}}_{\omega}^{\otimes N})^{\prime}$ results in diagonal matrix with the elements $(\alpha_{_{\emph{N}}} t)^{2}$.
Obviously, the largest eigenvalue $(N t)^{2}$ is doubly degenerate with the corresponding eigenstates
$|10^{\otimes N})\equiv |1\rangle\langle0|^{\otimes N}$ and $|01^{\otimes N})\equiv |0\rangle\langle1|^{\otimes N}$.
This leads one to prepare the initial state from the subspace $\{|0\rangle^{\otimes N},|1\rangle^{\otimes N}\}$  which results in  $\rho_{0}= |GHZ\rangle\langle GHZ|$ (as expected).


\section{Derivation of Eq. (\ref{3})}

In the case of an uncorrelated encoding process, $\Phi_{\omega}^{\otimes N}=\Pi_{i}\Phi_{\omega}^{(i)}$  ($\Phi_{\omega}^{(i)}$ denotes a parameter encoding map which acts on the $i$th probe), it is shown that
$(\Pi_{i}\tilde{\Phi}_{\omega}^{(i)})^{\prime \dag}(\Pi_{i}\tilde{\Phi}_{\omega}^{(i)})^{\prime}$ can be decomposed into two parts as

\begin{eqnarray}\label{sum1}
\nonumber & & \sum_{i=1}^{N}\tilde{\Phi}_{\omega}^{(1)\dagger}\tilde{\Phi}_{\omega}^{(1)}... \tilde{\Phi}_{\omega}^{(i)\prime\dagger}\tilde{\Phi}_{\omega}^{(i)\prime}...\tilde{\Phi}_{\omega}^{(N)\dagger}\tilde{\Phi}_{\omega}^{(N)}\\
\nonumber&+& \sum_{i\neq j=1}^{N}\tilde{\Phi}_{\omega}^{(1)\dagger}\tilde{\Phi}_{\omega}^{(1)}... \tilde{\Phi}_{\omega}^{(i)\prime\dagger}\tilde{\Phi}_{\omega}^{(i)}...\tilde{\Phi}_{\omega}^{(j)\dagger}\tilde{\Phi}_{\omega}^{(j)\prime}...\tilde{\Phi}_{\omega}^{(N)\dagger}\tilde{\Phi}_{\omega}^{(N)}.\\
\end{eqnarray}

The first summation involves the sum of $N$ terms whereas the second one involves the sum of $N(N-1)$ terms.
In the case of identical parameter encoding processes for $N$ probes, the above summations can be simplified to

\begin{equation}\label{sum2}
\sum_{i=1}^{N}\tilde{A}_{\omega}^{(1)}... \tilde{B}_{\omega}^{(i)}...\tilde{A}_{\omega}^{(N)}+\sum_{i\neq j=1}^{N}\tilde{A}_{\omega}^{(1)}... \tilde{C}_{\omega}^{(i)}...\tilde{C}_{\omega}^{(j)\dagger}...\tilde{A}_{\omega}^{(N)},
\end{equation}

where

\begin{eqnarray}
 \nonumber \tilde{A}_{\omega}^{(i)} &=& \mathcal{I}^{\otimes(i-1)}\otimes\tilde{\Phi}_{\omega}^{\dagger}\tilde{\Phi}_{\omega}\otimes \mathcal{I}^{\otimes(N-i)}, \\
 \nonumber \tilde{B}_{\omega}^{(i)} &=& \mathcal{I}^{\otimes(i-1)}\otimes\tilde{\Phi}_{\omega}^{\prime\dagger}\tilde{\Phi}_{\omega}^{\prime}\otimes \mathcal{I}^{\otimes(N-i)}, \\
 \tilde{C}_{\omega}^{(i)} &=& \mathcal{I}^{\otimes(i-1)}\otimes\tilde{\Phi}_{\omega}^{\prime\dagger}\tilde{\Phi}_{\omega}\otimes \mathcal{I}^{\otimes(N-i)},\\
 \nonumber
\end{eqnarray}

for $i=1,..,N$, and $\mathcal{I}$ is the identity matrix on the Liouville space $\mathcal{L}(\mathcal{H})$.

For the most general phase-covariant qubit map, $\Phi_{\omega}$, whose matrix form is

\begin{equation}\label{covariant map}
\tilde{\Phi}_{\omega} = \begin{pmatrix} J_{++} & 0 & 0 & J_{+-}\\ 0 & 0 & \eta_{\perp}e^{-i\phi} & 0\\ 0 & \eta_{\perp}e^{i\phi} & 0 & 0 \\ J_{--} & 0 & 0 & J_{-+} \end{pmatrix},
\end{equation}
one obtains

\begin{equation}\label{matrix1}
\nonumber \tilde{\Phi}_{\omega}^{\dagger}\tilde{\Phi}_{\omega}  = \begin{pmatrix} \frac{1+(k+\eta_{\parallel})^{2}}{2} & 0 & 0 & \frac{1+k^{2}-\eta_{\parallel}^{2}}{2}\\ 0 & \eta_{\perp}^{2} & 0 & 0\\ 0 & 0 & \eta_{\perp}^{2} & 0 \\ \frac{1+k^{2}-\eta_{\parallel}^{2}}{2} & 0 & 0 & \frac{1+(k-\eta_{\parallel})^{2}}{2} \end{pmatrix},
\end{equation}

\begin{equation}\label{matrix2}
\nonumber \tilde{\Phi}_{\omega}^{\prime\dagger}\tilde{\Phi}_{\omega}^{\prime}  = \begin{pmatrix} 0 & 0 & 0 & 0\\ 0 & t^{2}\eta_{\perp}^{2} & 0 & 0\\ 0 & 0 & t^{2}\eta_{\perp}^{2} & 0 \\ 0 & 0 & 0 &0 \end{pmatrix},
\end{equation}

\begin{equation}\label{matrix3}
\nonumber \tilde{\Phi}_{\omega}^{\prime\dagger}\tilde{\Phi}_{\omega}  = \begin{pmatrix} 0 & 0 & 0 & 0\\ 0 & -it\eta_{\perp}^{2} & 0 & 0\\ 0 & 0& it\eta_{\perp}^{2} & 0 \\ 0 & 0 & 0 &0 \end{pmatrix}.
\end{equation}

Considering the short-time expansion of the noise parameters as

\begin{eqnarray}
\nonumber \eta_{\perp} &=& 1-\alpha_{\perp}t^{\beta_{\perp}}+... \\
\nonumber \eta_{\parallel} &=& 1-\alpha_{\parallel}t^{\beta_{\parallel}}+...\\
  k &=& \alpha_{k}t^{\beta_{k}}+...,\\
  \nonumber
\end{eqnarray}

one finds that as long as $0\leq t<\tau$, the largest eigenvalue of
$(\Pi_{i}\tilde{\Phi}_{\omega}^{(i)})^{\prime \dag}(\Pi_{i}\tilde{\Phi}_{\omega}^{(i)})^{\prime}$ is $N^{2}t^{2}\eta_{\perp}^{2N}$,
and $\tau$ is the largest time that satisfies

\begin{equation}
\frac{2N^{2}}{(N-1)^{2}}\eta_{\perp}^{2}=1+k^{2}+\eta_{\parallel}^{2}+\sqrt{(1+k^{2}+\eta_{\parallel}^{2})^{2}-4\eta_{\parallel}^{2}}.
\end{equation}

In the case of unital channels which preserve identity, $k=0$, one obtains $\tau=(\alpha_{\perp} N)^{-1/\beta_{\perp}}$.

\section{Sufficient condition for obtaining the optimal measurement}

In quantum mechanics, a general measurement is mathematically represented by a collection of Hermitian positive semidefinite operators $\{E_{j}\}$ which satisfy $\sum_{j}E_{j}=\mathcal{I}$ and are named as positive operator-valued measures (POVM).
Since the probability of obtaining an experimental result $j$ is given by $tr(E_{j}\rho_{\omega})$, where $\omega$ is a specific value of the parameter, the classical version of $(\rho_{\omega}^{\prime}|\rho_{\omega}^{\prime})$ is written as

\begin{equation}\label{s4}
 F^{\downarrow}_{c}= \sum_{j}|\partial_{\omega}p(j|\omega)|^{2}=\sum_{j}|tr(E_{j}\rho_{\omega}^{\prime})|^{2}.
\end{equation}

Applying the Cauchy-Schwarz inequality to Eq.(\ref{s4}), one has

\begin{equation}
 F^{\downarrow}_{c} \leq \sum_{j} tr(E_{j})tr(E_{j}\rho_{\omega}^{\prime 2}).
\end{equation}

Equality in the Cauchy-Schwarz inequality is saturated if and only if $E_{j}^{1/2}\rho_{\omega}^{\prime}=\lambda_{j}E_{j}^{1/2}$ with $\lambda_{j}=tr(E_{j}\rho_{\omega}^{\prime})/tr(E_{j})$.
This shows that $E_{j}^{1/2}$ and, hence, $E_{j}$ are one-dimensional projectors onto the non-degenerate eigenspace of
$\rho_{\omega}^{\prime}$ corresponding to the eigenvalue $\lambda_{j}$.
For such measurements $F^{\downarrow}_{c}=tr(\rho_{\omega}^{\prime 2})$ and the saturability of Eq. (\ref{theorem}) in the main context is guaranteed.

\end{document}